%% file: main.tex
\documentclass[conference]{IEEEtran}
\IEEEoverridecommandlockouts
\usepackage{cite}
\usepackage{amsmath,amssymb,amsfonts}
\usepackage{algorithmic}
\usepackage{graphicx}
\usepackage{textcomp}
\usepackage{xcolor}
\usepackage{caption}
\usepackage{subcaption}
\usepackage{multirow}
\usepackage{booktabs}
\usepackage[breaklinks=true,colorlinks=true,linkcolor=black,urlcolor=black,citecolor=black]{hyperref}
\usepackage{cleveref}
    
\newcommand{\ignore}[1]{}

\newcommand\note[1]{\textcolor{red}{\textbf{#1}}}
    
\begin{document}

\captionsetup[figure]{labelfont={default},labelformat={default},labelsep=period,name={Fig.}}
\Crefname{figure}{Fig.}{Figs.}%

\title{\Large Patient Specific Biomechanics Are Clinically Significant In Accurate Computer Aided Surgical Image Guidance
\\[-1.0ex]
}


\author{Michael Barrow$^1$, Alice Chao$^1$, Qizhi He$^1$, Sonia Ramamoorthy$^2$, Claude Sirlin$^2$ and Ryan Kastner$^1$\\
$^1$ University of California San Diego $^2$ University of California San Diego School of Medicine\\ 
{\small mbarrow@eng.ucsd.edu, awchao@ucsd.edu, q9he@ucsd.edu, sramamoorthy@ucsd.edu, csirlin@ucsd.edu, kastner@ucsd.edu}
\\[-3.0ex]
}

\ignore{
\author{\IEEEauthorblockN{1\textsuperscript{st} Given Name Surname}
\IEEEauthorblockA{\textit{dept. name of organization (of Aff.)} \\
\textit{name of organization (of Aff.)}\\
City, Country \\
email address or ORCID}
\and
\IEEEauthorblockN{2\textsuperscript{nd} Given Name Surname}
\IEEEauthorblockA{\textit{dept. name of organization (of Aff.)} \\
\textit{name of organization (of Aff.)}\\
City, Country \\
email address or ORCID}
\and
\IEEEauthorblockN{3\textsuperscript{rd} Given Name Surname}
\IEEEauthorblockA{\textit{dept. name of organization (of Aff.)} \\
\textit{name of organization (of Aff.)}\\
City, Country \\
email address or ORCID}
\and
\IEEEauthorblockN{4\textsuperscript{th} Given Name Surname}
\IEEEauthorblockA{\textit{dept. name of organization (of Aff.)} \\
\textit{name of organization (of Aff.)}\\
City, Country \\
email address or ORCID}
\and
\IEEEauthorblockN{5\textsuperscript{th} Given Name Surname}
\IEEEauthorblockA{\textit{dept. name of organization (of Aff.)} \\
\textit{name of organization (of Aff.)}\\
City, Country \\
email address or ORCID}
\and
\IEEEauthorblockN{6\textsuperscript{th} Given Name Surname}
\IEEEauthorblockA{\textit{dept. name of organization (of Aff.)} \\
\textit{name of organization (of Aff.)}\\
City, Country \\
email address or ORCID}
}

}

\maketitle


\begin{abstract}

Augmented Reality is used in Image Guided surgery (AR IG) to fuse surgical landmarks from preoperative images into a video overlay. Physical simulation is essential to maintaining accurate position of the landmarks as surgery progresses and ensuring patient safety by avoiding accidental damage to vessels etc. 
In liver procedures, AR IG simulation accuracy is hampered by an inability to model stiffness variations unique to the patients disease.   
We introduce a novel method to account for patient specific stiffness variation based on Magnetic Resonance Elastography (MRE) data. 
To the best of our knowledge we are the first to demonstrate the use of in-vivo biomechanical data for AR IG landmark placement.
In this early work, a comparative evaluation of our MRE data driven simulation and the traditional method shows clinically significant differences in accuracy during landmark placement and motivates further animal model trials.  

\end{abstract}

\begin{IEEEkeywords}
Image Guided Surgery, Magnetic Resonance Elastography, Physical Simulation, Augmented Reality
\end{IEEEkeywords}


\input{texsrc/introduction}
\input{texsrc/method}
\input{texsrc/evaluation} 
\input{texsrc/conclusion} 


\bibliography{bibliography.bib}

\end{document}

%% file: texsrc/introduction.tex
\section{INTRODUCTION}

Primary liver cancer has the fastest growth of incidence and the second highest mortality of all cancers in the United States~\cite{siegel2019cancer}. It is estimated that over one million people will die from liver cancer in 2030 \cite{whoprojections}. 
 A partial hepatectomy is a common treatment to remove cancerous primary and metastatic lesions where the diseased parts of the liver are excised. A crucial part of the operation is understanding where the tumors, vessels, and other important structures/landmarks are located. To aid in this, the patient typically undergoes preoperative cross-sectional imaging (e.g., CT/MRI scans). Surgeons use these images to determine resectability based upon the location of important structures (e.g., portal pedicles and hepatic veins),  analyze tumor margins, accurately compute future liver remnant volumes, and generally aid in surgical planning and navigation. However, it can be challenging to register preoperative cross-sectional images to the surface of the liver at the time of operation since surgical actions cause significant and sometimes permanent liver deformations that lead to mismatches with cross-sectional images~\cite{heizmann2010assessment}. Furthermore, the liver can be mobilized further distorting anatomic relationships and liver shape. Mentally integrating preoperative data into the operative field is time consuming and error prone ~\cite{hansen2010illustrative}. This can make it difficult to accurately localize smaller tumors intra-operatively, which can affect surgical decision making and adequate resection of primary and metastatic liver tumors.

Computer-assisted image guidance aids surgeons by enhancing surgical video feed with added guidance landmarks. In liver procedures, the goal is to assist the surgeon with error prone manual image mappings. 
Augmented Reality (AR) Image Guidance (IG) is a mapping method that fuses preoperative scans with intraoperative images to provide more detailed information about the surgical site. Typically AR IG merges the preoperative CT/MRI data directly into the surgeons view~\cite{bernhardt2017status}. %
Accurate real-time image guidance is a valuable surgical tool that leads to more precise surgical procedures. The state of the art in computer assisted image guidance for hepatic surgery does not use patient specific biomechanical models to position surgical landmarks in 
AR image guidance~\cite{plantefeve2016patient,haouchine2015impact,soler2008patient,soler2014real}.  
We propose a data driven approach that uses patient specific liver mechanics to estimate the landmark positions. Magnetic Resonance Elastography (MRE) provides an accurate method to measure tissue mechanical properties \cite{xanthakos2014use,shi2014short,zhang2016short,yasar2016interplatform,hines2010repeatability,venkatesh2014magnetic}. Moreover, it is non-invasive, and it can be acquired in conjunction with MRI images which are commonly used in conventional image guided liver operations. We believe that we are the first to propose the use of these measurements for image guided hepatic surgery. 
 
Our initial studies described in this manuscript indicate that modeling using MRE based physical simulation versus modeling with a constant value ``atlas'' (i.e., current state of the art) results in clinically significant differences in IG accuracy. 
Overall, while demonstrating the improvement in AR IG accuracy with our approach, \textbf{we make the following three contributions:}\\

\noindent \textbf{Firstly; The need for patient specific biomechanics is demonstrated}. We analyzed the MRE data from a cohort of 120 patient scans and found liver stiffness commonly assumed in AR IG is unrealistic.\\ 
\textbf{Secondly; Clinically significant AR IG accuracy improvement is demonstrated}. Using the MRI and MRE data of our cohort, we simulated liver retraction procedures. The simulations show that MRE based image guidance more accurately models the patient's liver response to a surgical tool interaction when compared to conventional techniques.\\ 
\textbf{Thirdly; Our method is validated using convergence of our simulation with theory}. We construct a liver phantom model using parameters from characterized silicone. We compare the theoretical Euler-Bernoulli response of the phantom under a retraction load to the response of our MRE method. The convergence study found our method to be sound.\\

Taken together these three studies motivate, test and demonstrate the hypothesis that patient specific biomechanics from MRE can improve AR IG in liver procedures. 

\begin{figure*}[!htb]

    \includegraphics[width=1.0\textwidth,trim={0cm 0cm 0cm 0cm},clip]{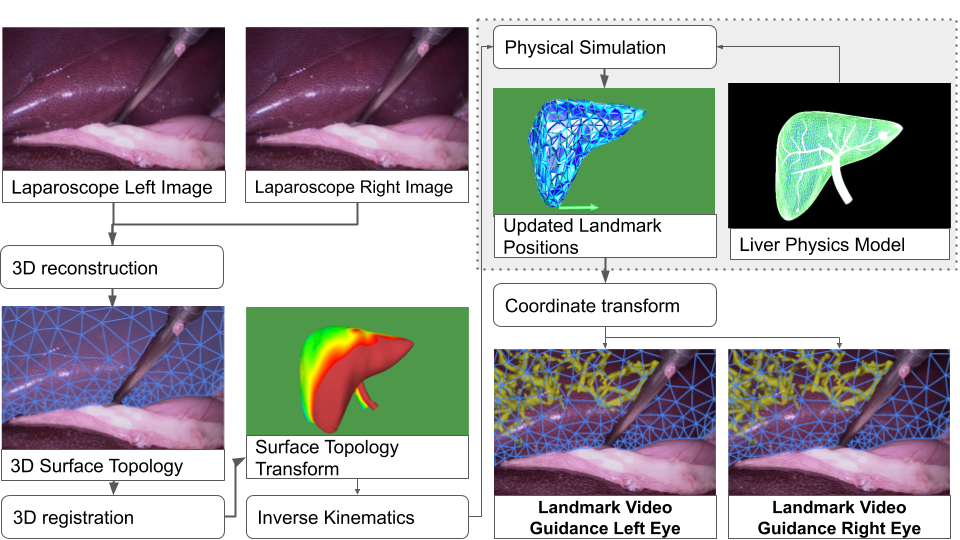}%

    \caption{Block diagram of an AR IG pipeline. For each video frame; Firstly, the 2D laparoscope video is converted to 3D. Secondly, registration maps the visible portion of liver to a complete 3D surface model. Thirdly, changes on the 3D surface are propagated inside the liver using a physical simulation. Finally, updated guidance landmark positions are displayed on a modified laparoscope video feed. This work focuses on accurate physical simulation (gray area) which is key to accurate IG.
    \\[-5.0ex]
    }\label{tab:concept}
    
\end{figure*}

%% file: texsrc/method.tex
\section{METHODS}\label{sec:algorithm}

We performed three experiments to better understand the effect of MRE based physical simulation on augmented reality image guidance accuracy. 
Firstly, we use MRE scans of a patient cohort to check if traditional AR IG stiffness assumptions are reasonable.
Secondly, we study how our method can improve accuracy by comparing AR IG landmark placement between MRE based and traditional IG simulations.
Thirdly, we use established theory to validate our simulation method with a convergence study.

\subsection{Study of Variations in Patient Liver Stiffness}\label{sec:CohortStudy}

There is little literature on the variation and distribution of liver stiffness because of the difficulty of measuring it in vivo. We use a statistical study of liver stiffness measured using MRE to determine if variations of stiffness in a sample population motivate using MRE data for patient specific image guidance. It would be interesting to use MRE data in AR IG if a significant portion of the cohort shows a large difference in liver stiffness from a atlas value of 2.1kPa commonly used in AR IG guidance simulation~\cite{lee2013mr}.


\begin{figure}[!htb]
    \includegraphics[width=0.48\textwidth,trim={0cm 0cm 6cm 3cm },clip]{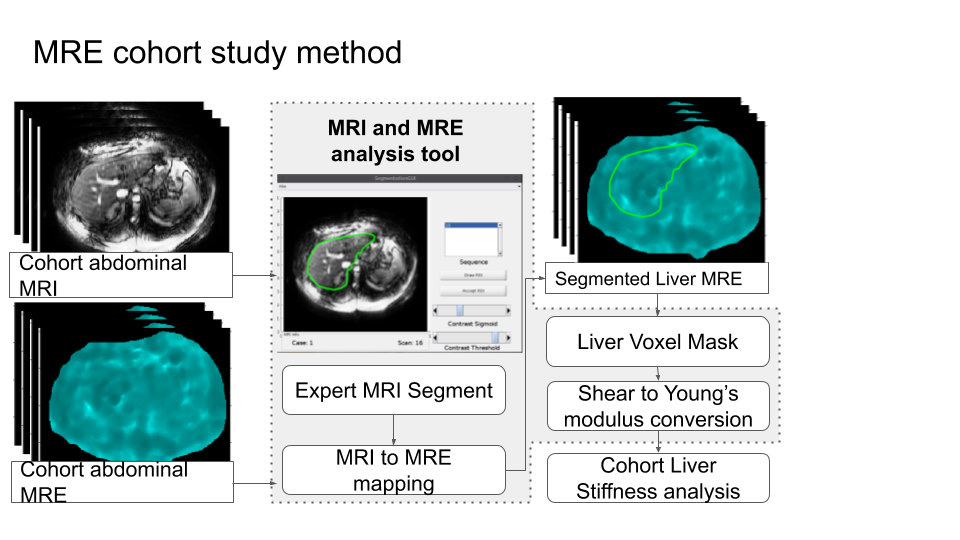}%
    \caption{Major steps of our liver stiffness case study. Using our tool (enclosed in gray); Firstly, liver was segmented from the cohort abdominal MRI. Secondly, liver was segmented from MRE using a MRI voxel mapping. Thirdly, the Young's modulus of the liver was calculated.
    \\[-6.0ex]
    }\label{fig:Segment:MRI}
    \label{fig:MeanCohortStiffness:Tool} 

\end{figure}

We used 120 2D MRE scans acquired between March 2012 and December 2013 under a Health Insurance Portability and Accountability Act (HIPAA)-compliant study. Our cohort sample population was adults (\textgreater 18 years of age) who were severely obese (body mass index [BMI] \textgreater 35kg/m$^2$) and were being evaluated for weight loss surgery. The exclusion criteria were contraindications to MRI or history of known liver disease besides potential nonalcoholic fatty liver disease.
MRE were acquired using a GE Medical Systems Discovery\texttrademark MR750 system at 3.0T field strength. Voxel resolution was 1.64$\times$1.64mm, slice thickness was 10mm and space slicing was 10mm. 44ml MultiHance contrast bolus agent was administered to patients by IV route.

We created a tool to segment the liver portion of 2D MRE DICOM files and extract liver stiffness. Major steps of the MRE analysis process are shown in~\cref{fig:MeanCohortStiffness:Tool}. The tool allowed a imaging expert to define a ROI polygon of the liver on top of the MRI sequence used to create MRE elastograms. Once the expert had defined a ROI polygon, our tool mapped the polygon to a corresponding MRE. 

After the cohort MRE segmentation, we masked away stiffness voxels outside of the liver ROI using a binary inclusion test. Next the mean shear modulus is computed on the masked MRE. Then, shear modulus is converted to a Young's modulus. Finally we create a histogram of liver stiffness from the segmented cohort data set.
%
%

\subsection{Study of MRE Significance in Image Guidance}\label{sec:AtlasEFGandMeanEFGCompare}

Accidental damage to vessels in the liver can complicate surgery. Our MRE based AR IG method could improve patient safety if clinically significant difference exists between blood vessel landmark positions using our method and a traditional approach. We studied the difference of blood vessel landmark placement in a liver retraction procedure, a common step in liver surgery. 

\begin{figure}[!ht]
        \includegraphics[width=0.48\textwidth,trim={2cm 0cm 3.5cm 2.5cm },clip]{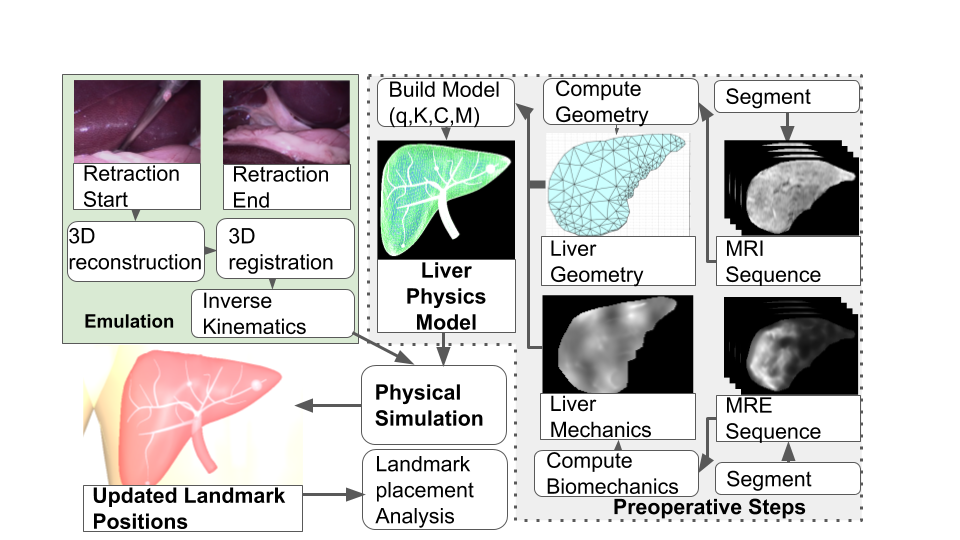}%
    \caption{Block diagram of liver retraction used to evaluate clinical significance of our novel MRE data driven AR IG approach. \emph{To the best of our knowledge, our physical model is unique in integrating in-vivo measurements of stiffness}. Our simulation procedure was applied to 113 of our 120 cohort using their MRE data. Firstly in the offline steps; A liver physics model was created from MRI and MRE scans (gray region). Secondly a hoisting load was computed using the physics model (green region). Next the retraction is simulated in an online stage and IG landmarks are positioned. Finally our blood vessel positions are compared for clinically significant difference with traditional AR IG in a post processing analysis.
    }
    \label{fig:RetractionExperiment:Experiment} 
\end{figure}

~\Cref{fig:RetractionExperiment:Experiment} shows how we adapted the SOFA AR IG framework 
to implement our MRE method. We compared retraction studies of our cohort simulated using our method to simulations using a traditional AR IG approach. 

\subsubsection*{MRE Data Driven Physical Simulation}

Our patient specific AR IG method uses a mesh free physical simulation. Mesh free approaches are suited to AR IG because they do not require re-meshing after topology changes (such as cutting) that occur during a resection and have been previously implemented in SOFA~\cite{faure2011sparse}. Our MRE simulation uses the following canonical semi-discrete formulation:
\begin{equation}
    \mathbf{ M\ddot{q} - f_{int}(q,\dot{q}) = f_{ext}(q,\dot{q})}
\end{equation}
    
where $\mathbf{q}$ is the nodal displacement vector, $\mathbf{\dot{q}}$ is the nodal velocity vector, $\mathbf{\ddot{q}}$ is the nodal acceleration vector, $\mathbf{M}$ is the mass matrix, $\mathbf{f_{int}}$ is the internal force vector, and $\mathbf{f_{ext}}$ is the external force vector. 
    
We choose a computationally efficient discretization of this system  by adopting the \textit{Implicit Euler} method~\cite{baraff1998large} which will create a system of linear equations. This is because our cohort MRE data encodes linear mechanics and because linear approximations are used to achieve video rate performance in AR IG. The system of equations is fed into the Conjugate Gradient (CG) descent method that will ultimately solve for $\delta {\mathbf{\dot{q}}}$ for each time step. The implicit Euler method is adopted as follows:
\begin{equation}\label{eqn:ImplicitEulerMethod}
         (\mathbf{M}-h \mathbf{C}-h^2 \mathbf{K}) \delta \mathbf{\dot{q}} = h (\mathbf{f_{ext}}+h \mathbf{K} \mathbf{\dot{q}}) 
\end{equation}

where $ h $ is the time step (this is in milliseconds for our study), $\mathbf{K}$ is the Jacobian of first order derivatives of $\mathbf{f_{int}}$ with respect to $\mathbf{\dot{q}}$ (also known as the stiffness matrix), and $\mathbf{C}$ is the Jacobian of first order derivatives of $\mathbf{f_{int}}$ with respect to $\mathbf{q}$ (also known as the damping matrix). Note that $K$, $C$ and $M$ may be precomputed offline and that this corresponds to ``building the simulator model'' from processed preoperative MRI and MRE in~\cref{fig:RetractionExperiment:Experiment}.

The next step of~\cref{fig:RetractionExperiment:Experiment}, Physical Simulation, is online.~\Cref{eqn:ImplicitEulerMethod} is solved numerically via the following mappings:

\begin{equation}
    {A = (\mathbf{M} - h\mathbf{C} - h^2\mathbf{K})}
\end{equation}
\begin{equation}
    {\mathbf{b} = h(\mathbf{f_{ext}} + h\mathbf{K\dot{q}})}
\end{equation}    
\begin{equation}
    {\mathbf{x} = \delta \mathbf{\dot{q}}}
\end{equation}    

Such that $A\mathbf{x} = \mathbf{b}$, where $\mathbf{x}$ contains the vector encoding of surgical landmark positions to resolve for the current time step. 
Once $\delta \mathbf{\dot{q}}$ is solved for each node in the system, calculating the change in global position, $\delta \mathbf{q}$, is simply done by integration.

Our MRE method can improve AR IG accuracy with no computational overhead since it can be shown that online computation method of $\delta \mathbf{\dot{q}}$ is the bottleneck. First, there is an initial guess: $\mathbf{x_0}$
\begin{equation}
 {\mathbf{r_n} = \mathbf{b} - A\mathbf{x_n}}
\end{equation}
\begin{equation}
 {{\alpha}_n = \dfrac{\mathbf{r_n^T}\mathbf{r_n}}{\mathbf{p_n^T} A \mathbf{p_n}}}
 \end{equation}
\begin{equation}
{\mathbf{x_{n+1}} = \mathbf{x_n} + {\alpha}_n \mathbf{p_n}}
\end{equation}
\begin{equation}
{0 \leq n \leq N, n \in \mathbf{Z}}
\end{equation}

\centerline{(When $n = 0$, $\mathbf{p_0}$ is initially set to $\mathbf{r_0}$)}

The steps above illustrate a portion of one iteration of the conjugate gradient descent method, with each subsequent $\mathbf{x_n}$ representing a closer approximation of $\mathbf{x}$. The linear solver continues to iterate until the loop is terminated upon reaching preset error bounds (in the example given, this happens when $n = N$ which is 200 in our study). Since MRE data is used to construct $A$ it does not exacerbate the CG bottleneck.

\subsubsection*{MRE Liver Physics Model}

We create a SOFA~\cite{allard2007sofa} Physics Model ``node'' that can be integrated into any SOFA AR IG system. The node is composed of a stiffness matrix ($\mathbf{K}$) created from MRE data, geometry created from MRI data, a mass matrix ($\mathbf{M}$) computed using MRE and MRI data and a displacement vector $\mathbf{q}$ that is initialized from MRI data and updated by the physical simulation during AR IG. 

Patient liver models are built using the procedure depicted in~\cref{fig:RetractionExperiment:Experiment} (gray region). Firstly liver data from our cohort MRI and MRE scans is segmented using the tool described in~\cref{fig:MeanCohortStiffness:Tool}. Next surface geometry is created from MRI data using a modified Delaunay triangulation~\cite{engwirda2014locally}. Liver biomechanics are then extracted from MRE data. We first convert voxel intensity to elastic modulus, then sub-sample this for real time AR IG using a 3D Voronoi decomposition of the volume to choose DOFs. After $\mathbf{M}$ is computed, $\mathbf{K}$,$\mathbf{M}$, and $\mathbf{q}$ are assembled into the SOFA node.

\subsubsection*{Clinical Significance Study Methodology}

\Cref{fig:RetractionExperiment:Experiment} shows the major steps of a retraction experiment using our MRE AR IG method. We compare blood vessel positions from our method with a traditional AR IG method. Our focus is on clinical significance of our landmark placement method and our experimental pipeline differs from the canonical AR IG pipeline in~\cref{tab:concept}. We emulate AR steps prior to physical simulation (\cref{fig:RetractionExperiment:Experiment} green region) because accurate real time implementations of all major AR steps remain open research problems. Laparoscopic video input is noisy and many accuracy trade offs are made in the 3D reconstruction and registration steps that would make analysis of our Physical simulation method difficult. Liver retraction (hoisting) video input is therefore emulated using a virtual implementation of the 10mm diameter Nathanson retractor shown hoisting the liver in~\cref{fig:RetractionExperiment:Experiment}. 

The traditional AR IG approach is an implementation of Plantafeve's state of the art model~\cite{plantefeve2016patient} that uses a traditional ``atlas'' reference stiffness from Lee's study~\cite{lee2013mr}. Traditional AR IG can be directly compared with our data driven simulations by using the same emulated retraction input as our MRE method.

Liver retractions were simulated using the same cohort described in~\cref{sec:CohortStudy}.
We assumed the hoisting force was equal to the gravitational force of the liver and was coplanar on the transverse plane, with the Nathanson retractor force applied to the lower inferior right lobe. We further assume a linear elastic resistance from the abdomen and that friction is negligible. We rejected 17 of our cohort scans where it was not possible to simulate a retraction because of bad geometry, leaving 103 cases of vessel placement compared.


\subsection{Validation of Our Image Guidance Approach}\label{sec:FEAandEFGCompare}

It is extremely challenging to validate AR IG physical simulation in vivo. Difficulties center on the problem of obtaining a "ground truth" of the initial and steady state of the surgical scene owing to a lack of accurate procedurally approved measuring tools. However there a host of other complications including the ethical questions of destructive material characterizations on live subjects, unknown contact coefficients, and unknown boundary conditions.

\subsubsection*{Physics Simulator Theoretical Convergence Study}

Standard AR IG validation approach is to measure simulation registration error with a phantom (tissue proxy) or a theoretical model~\cite{haouchine2013image,haouchine2014towards}. Because MRE validation has been reported using theoretical models~\cite{hollis2016computational} we choose a theoretical validation. We simulate a cantilever beam under distributed load and check for convergence of our simulation method with an established theoretical model of the system and our mesh free model. 

\begin{figure}[!h]

\includegraphics[width=0.48\textwidth,trim={2cm 0cm 5cm  4cm},clip]{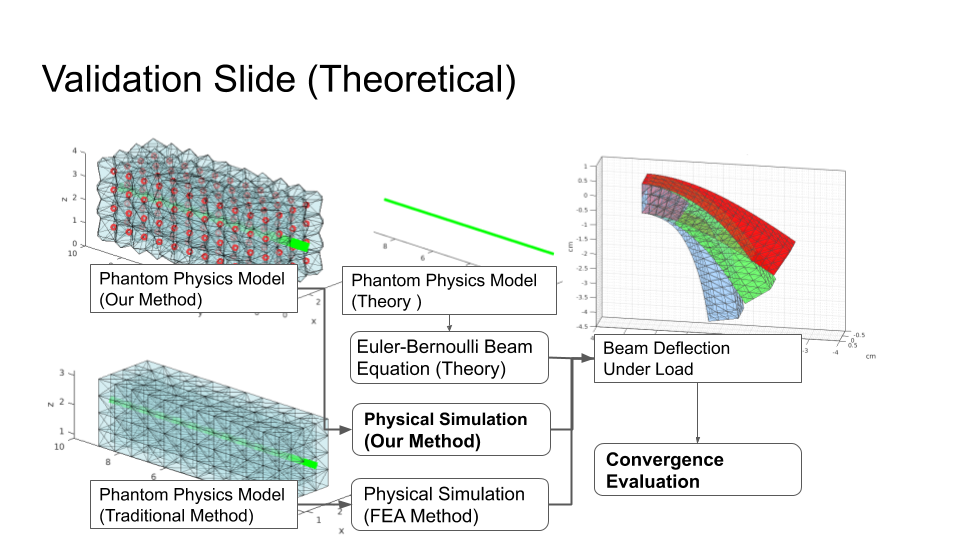}
\caption{Steps in our IG landmark placement accuracy study. Our landmark placement physical simulation method is compared with a traditional baseline and a theoretical model. Because simulations that converge to the theoretical model are accurate, beam deflection error can be used to validate our MRE based landmark positioning. Comparing our error with error of the traditional method benchmarks our performance against state of the art.}\label{fig:validation:Experiment}

\end{figure}

In addition to testing our simulation method, we use a conventional Finite Element Analysis (FEA) simulation method to model beam deflection as a convergence baseline. This helps better understand the accuracy and speed trade off of our real time mesh free model.

Major steps of our validation study are shown in~\cref{fig:validation:Experiment}. Our simulation method and the traditional method are implemented using the SOFA framework. The theoretical model is a closed form Euler-Bernoulli model and beam deflection can be computed directly. 
All three beams model a silicone phantom since the phantom stiffness can be reasonably linearized for a validation model~\cite{liu2013analytical}, our MRE scans have linear elastic modulus and beam models have been used for SOFA mesh free validation in the literature~\cite{faure2011sparse}. We choose a high resolution FEA method with 1.64mm$^3$ element geometry as the baseline. 

We use a shape function to compare our mesh free beam to the theoretical beam as they have different coordinate systems.  
The mapping result can be seen in the ``Phantom Physics Model (Our Method)'' of~\cref{fig:validation:Experiment}. The green line is the Voronoi to Euler-Bernoulli coordinate mapping. We then use registration distance to compute the convergence error as is common in the literature~\cite{haouchine2013image,haouchine2014towards}. 
Details of the shape function can be found in~\cite{faure2011sparse}. The procedure for the FEA baseline is similar except FEA DOFs coincident with the green line of the ``Phantom Physics Model (Traditional Method)'' in~\cref{fig:validation:Experiment} can be directly compared with the theoretical model. The Euler-Bernoulli beam equation reads:
\begin{equation}
w(x)=\frac{f_{load}x^2(6L^2-4Lx+x^2)}{24EI} 
\end{equation}

where $w(x)$ is displacement of the central beam axis in the load direction, $E$ is Young's modulus, $L$ is the long edge and $I$ is the second moment of beam cross sectional area.

\subsubsection*{Simulation Validation Configuration}
For validation simulation geometry, a hyperrectangle beam with idealized axisymmetric AAA geometry was  discretized with dimensions \textit{l}=50mm, \textit{w}=10mm, \textit{h}=10mm. $E$ was the 12kPa Young's modulus of a well characterized 10\% silicone phantom from Clear Ballistics LLC. 
We set loading and convergence criteria following Plantefeve's patient specific modeling approach for liver vessel tracking~\cite{plantefeve2016patient}. 

%% file: texsrc/evaluation.tex
\section{RESULTS AND DISCUSSION}\label{sec:architecture}

Overall the results of our studies in~\cref{sec:algorithm} showed that our MRE AR IG method was more accurate than the traditional approach. Although we focused on introducing a novel physical simulation and liver modeling approach in this manuscript, our method plugs in to the existing SOFA AR IG framework and runs at video frame rate.

\subsection{Variations in Patient Liver Stiffness}

\begin{figure}[!h]
    \includegraphics[width=0.44\textwidth]{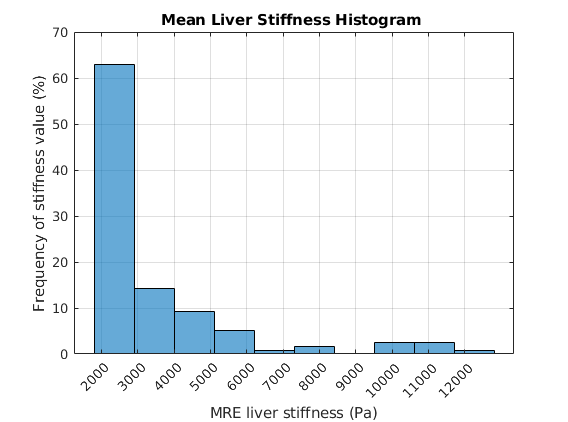}
      \caption{Histogram of mean liver stiffness. A large fraction of the MRE scans are significantly stiffer than a 2.1kPa atlas value which motivates using MRE for AR simulation.}\label{fig:MeanCohortStiffness:histogram}
\end{figure}

 Our \emph{study of variations in patient liver stiffness} in~\cref{sec:CohortStudy} motivated using MRE data in AR IG since traditional assumptions about liver stiffness were found unreasonable.~\Cref{fig:MeanCohortStiffness:histogram} is the histogram of cohort liver stiffness generated by running the liver MRE analysis procedure illustrated by~\cref{fig:MeanCohortStiffness:Tool}. Of the 120 liver scans analyzed, 30\% were over 1kPa stiffer than the ``atlas'' value of 2.1kPa used in the state of the art~\cite{lee2013mr}. Furthermore, 20\% were more than twice as stiff. Because of the large stiffness variation and since liver stiffness typically increases as disease progresses~\cite{yoneda2008noninvasive,mueller2010liver}, we demonstrated that it is interesting to consider MRE stiffness data in AR IG for liver procedures.

\subsection{MRE Significance in Image Guidance Accuracy}\label{sec:architecture:accuracy}

Our \emph{study of MRE significance in image guidance} in~\cref{sec:AtlasEFGandMeanEFGCompare} demonstrated that using MRE patient data in physical simulation could provide more accurate AR IG than traditional methods. We studied blood vessel landmark placement of our method (illustrated in~\cref{fig:RetractionExperiment:Experiment}) and compared it to a traditional AR IG method~\cite{plantefeve2016patient} in 103 cases. 
5mm landmark placement difference was considered clinically significant~\cite{bernhardt2017status}.

\begin{figure}[!h]
    \centering
    \begin{subfigure}{0.24\textwidth}        
    \includegraphics[width=1\textwidth,trim={0 0 7cm 0},clip]{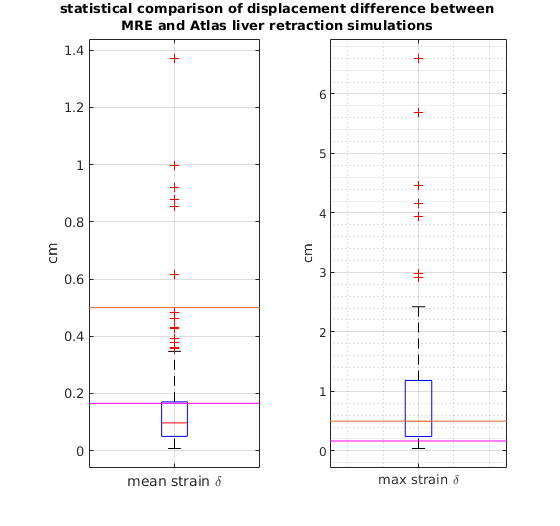}%
    \vspace*{-5mm}\caption*{   (i)}\label{fig:strainboxplot:mean}
    \end{subfigure}
    \hspace{-0.4em}
    \begin{subfigure}{0.24\textwidth}        
    \includegraphics[width=1\textwidth,trim={7cm 0 0 0},clip]{images/evaluation/ComparisonofDisplacementMREandAtlasLiverGeometryBoxPlot}\hspace{-8em}%
    \vspace*{-5mm}\caption*{(ii)}\label{fig:strainboxplot:max}
    \end{subfigure}

      \caption{Box plot showing liver IG difference of our method vs the traditional from our 103 patient case study.~\Cref{fig:strainboxplot}i shows most cases agree to within one MRI voxel (magenta line) except near surgical tools.~\Cref{fig:strainboxplot}ii shows IG error is often clinically significant (orange) near surgical tools where accuracy is most critical.
      }
      \label{fig:strainboxplot} 

\end{figure}

~\Cref{fig:strainboxplot}i plots the mean displacement difference \emph{throughout the whole volume of liver} for each pair of retractions compared. Averaging over the whole liver, the two methods agree to within a 1.64mm MRI voxel (magenta line).~\Cref{fig:strainboxplot}ii plots the displacement difference \emph{at the retractor}. There is clinically significant difference in accuracy (orange line) between our method and the traditional at the tool for 50\% of cases.

%
%

\begin{figure*}[!h]
\begin{tabular}{cc}
\begin{subfigure}{0.5\textwidth}\includegraphics[width=1\textwidth,trim={0cm 0cm 0cm 0cm },clip]{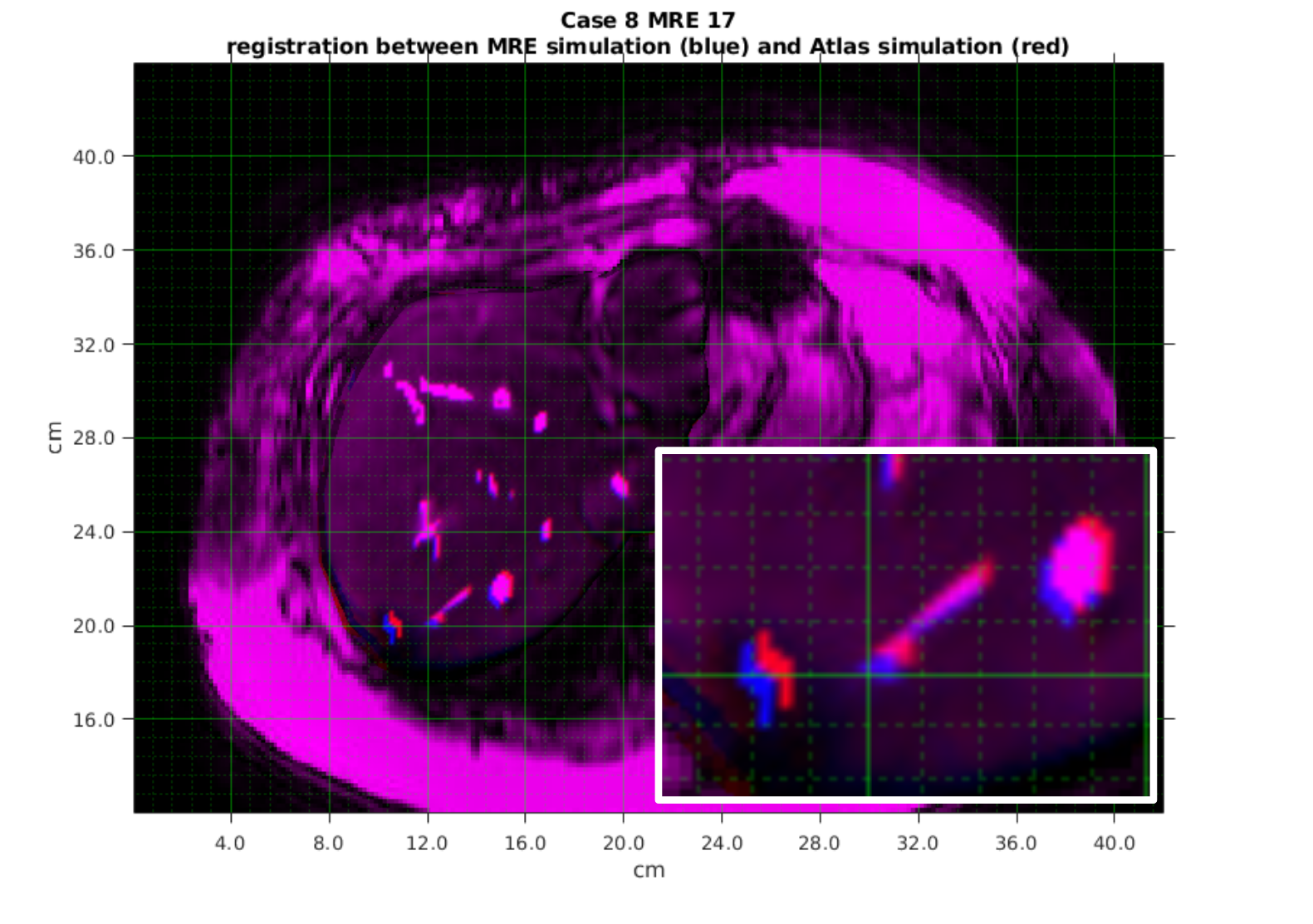}\vspace*{-5mm}\caption{Mean strain $\delta$ with error zoom}\label{fig:guidance:meanerr}\end{subfigure}
 &  
\begin{subfigure}{0.5\textwidth}\includegraphics[width=1\textwidth,trim={0cm 0cm 0cm 0cm },clip]{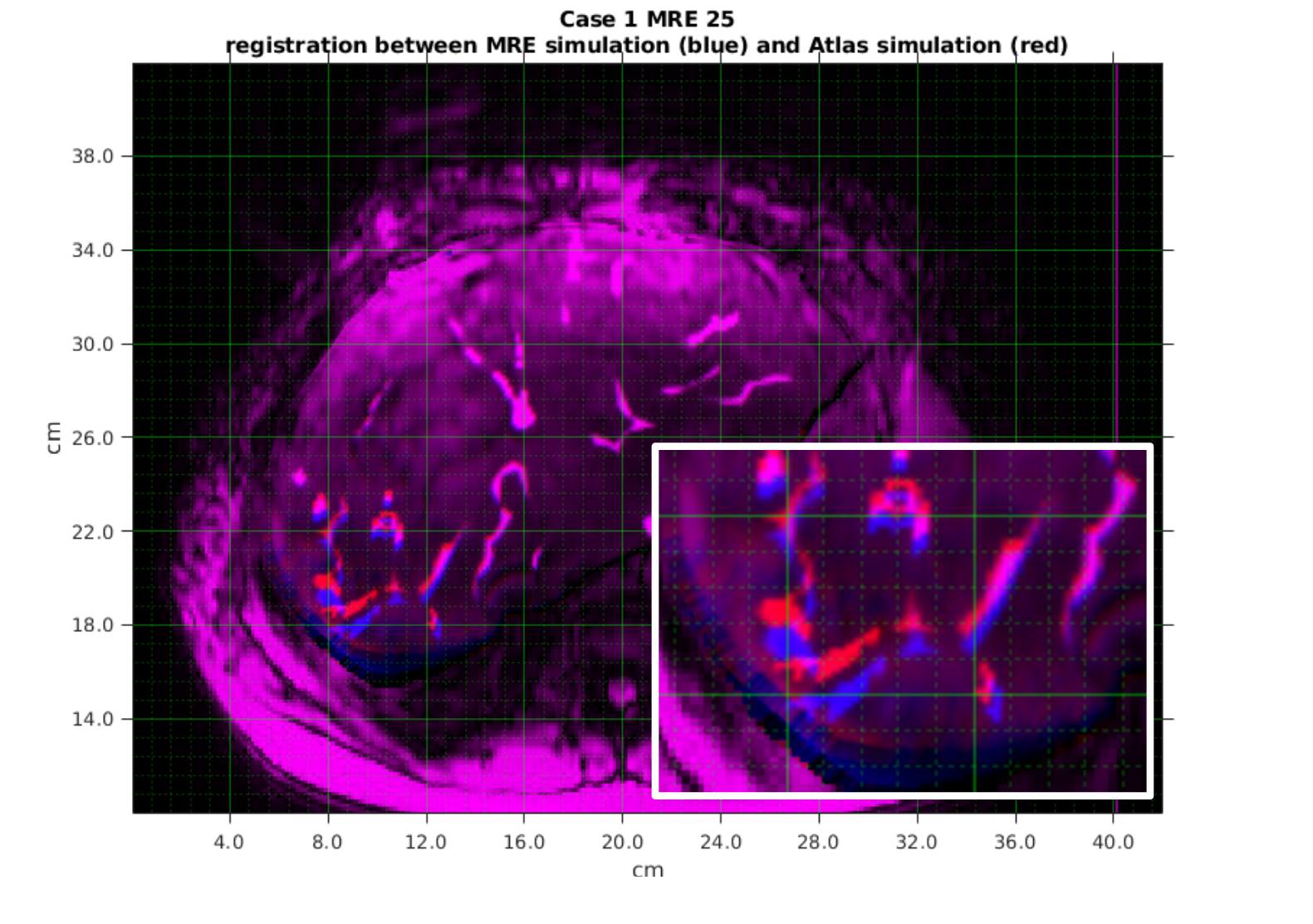}\vspace*{-5mm}\caption{+1 standard deviation strain $\delta$ with error zoom}\label{fig:guidance:sd1err}\end{subfigure}  
\\ 
\begin{subfigure}{0.5\textwidth}\includegraphics[width=1\textwidth,trim={0cm 0cm 0cm 0cm },clip]{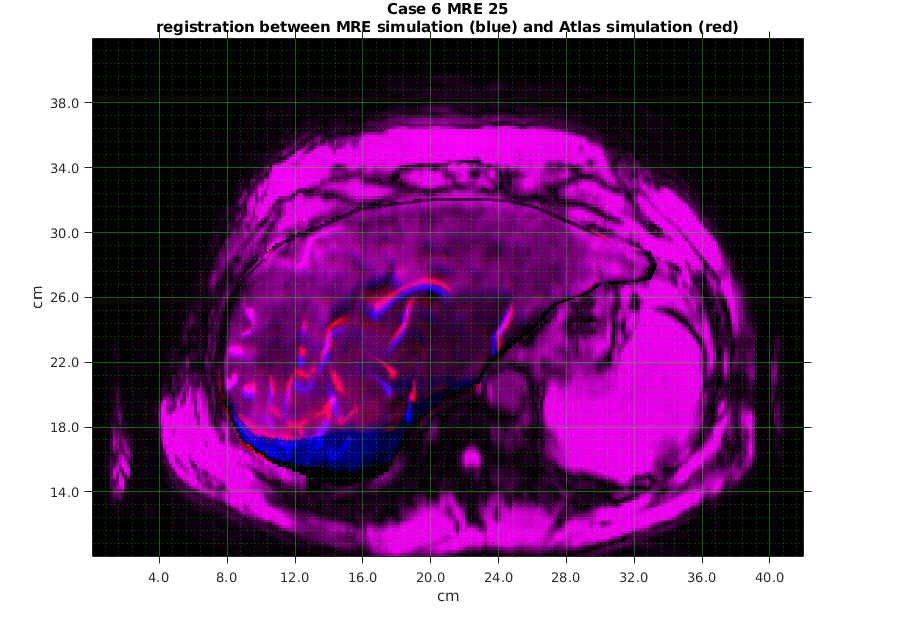}\vspace*{-5mm}\caption{+2 standard deviation strain $\delta$}\label{fig:guidance:sd2err}\end{subfigure}
& 
\begin{subfigure}{0.5\textwidth}\includegraphics[width=1\textwidth,trim={0cm 0cm 0cm 0cm },clip]{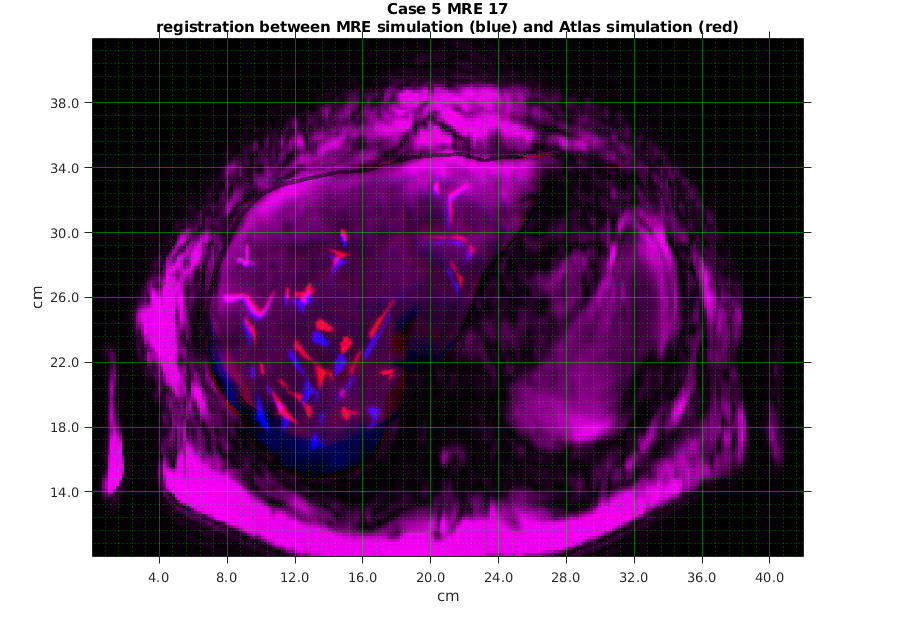}\vspace*{-5mm}\caption{strain $\delta$ outlier}\label{fig:guidance:outlier}\end{subfigure}
\end{tabular}

\caption{Simulations showing effects of incorrect stiffness on guidance. The MRE and atlas retraction simulations are blue and red channels in each image, respectively. Differences in image color indicate where features are misplaced. The green grid shows the scale of error. Mean strain $\delta$ has vessel misplacement concentrated around the retractor load. As strain $\delta$ increases, vessel misplacement also increasingly spreads throughout the liver.
\\[-5.0ex]
}\label{fig:guidance}

\end{figure*}

%
%

\Cref{fig:guidance} shows the clinical implications of~\cref{fig:strainboxplot} by illustrating retraction comparisons for standard deviations of~\cref{fig:strainboxplot}ii. Where the two simulation methods agree the images appear purple. Where they disagree, the image appears more blue or red. It is clear from the contrast enhanced vessels in~\cref{fig:guidance} that the difference of landmark placement in the traditional simulation method would impact a surgeon's ability to effectively use AR IG for vessel avoidance.

\subsection{Validation of Our Image Guidance Approach}

A \emph{Physics Simulator Theoretical Convergence Study} in~\cref{sec:FEAandEFGCompare} validated the accuracy of our mesh free AR IG landmark placement method by checking its convergence with Euler-Bernoulli beam theory. In addition, a traditional AR IG physical simulation had comparable theoretical convergence as compared to our method.

\begin{figure}[!h]
\includegraphics[width=0.48\textwidth]{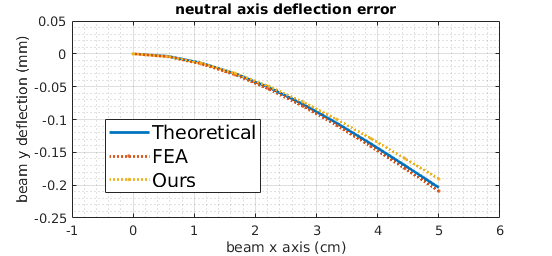}\caption{Convergence evaluation of Mesh Free (ours) and FEA simulation convergence with theory for our retraction study (\cref{fig:RetractionExperiment:Experiment}). Both our method and traditional FEA simulations agree with theory to within 1\% of the 1.64mm MRI resolution although FEA is more accurate. Our simulator accuracy is acceptable since AR IG landmarks are at minimum 1 voxel.}\label{fig:validation:DeflectionError}
\end{figure}

~\Cref{fig:validation:Experiment} illustrates the validation procedure and~\cref{fig:validation:DeflectionError} shows the convergence error for the retraction experiments run on our cohort (\cref{sec:AtlasEFGandMeanEFGCompare}). Both our method and the traditional method converge to within 1\% of the 1.64mm MRI resolution. This convergence is sufficient to accurately model the positioning of a 1 voxel IG landmark in a liver retraction. Although the traditional FEA physics simulation method more closely approximates Euler-Bernoulli theory, our method can be far more accurate in practice by using MRE data to account for patient specific biomechanics as shown in~\cref{sec:architecture:accuracy}.

%% file: texsrc/conclusion.tex
\section{CONCLUSIONS}\label{sec:conclusion}

We demonstrated a novel method of surgical landmark placement for augmented reality image guidance in liver surgery.
Our method is the first to apply patient specific biomechanics from magnetic resonance elastography to the problem of accurately intraoperatively tracking surgical landmarks within the liver using a physical simulation.

We used three studies to motivate, test, and validate our image guidance approach. Our results indicate our simulation method could improve image guidance accuracy by clinically significant margins.
Although we focus on introducing the use of patient elastogram scans to enhance surgical landmark placement, our method is plug in compatible with the SOFA framework which has been successfully applied to augmented reality image guidance in liver procedures.
In conclusion, our promising preliminary results motivate animal model trials where our landmark placement method may be integrated into a SOFA based AR IG system.